\PassOptionsToPackage{unicode}{hyperref}
\PassOptionsToPackage{hyphens}{url}
\PassOptionsToPackage{dvipsnames,svgnames,x11names}{xcolor}
\documentclass[
  10pt,
]{article}
\usepackage{xcolor}
\usepackage[margin=2.5cm]{geometry}
\usepackage{amsmath,amssymb} 
\usepackage{iftex}
\ifPDFTeX
  \usepackage[T1]{fontenc}
  \usepackage[utf8]{inputenc}
  \usepackage{textcomp} % provide euro and other symbols
\else % if luatex or xetex
  \usepackage{unicode-math} % this also loads fontspec
  \defaultfontfeatures{Scale=MatchLowercase}
  \defaultfontfeatures[\rmfamily]{Ligatures=TeX,Scale=1}
\fi
\usepackage{lmodern}
\ifPDFTeX\else
\fi
\IfFileExists{upquote.sty}{\usepackage{upquote}}{}
\IfFileExists{microtype.sty}{% use microtype if available
  \usepackage[]{microtype}
  \UseMicrotypeSet[protrusion]{basicmath} % disable protrusion for tt fonts
}{}
\makeatletter
\@ifundefined{KOMAClassName}{% if non-KOMA class
  \IfFileExists{parskip.sty}{%
    \usepackage{parskip}
  }{% else
    \setlength{\parindent}{0pt}
    \setlength{\parskip}{6pt plus 2pt minus 1pt}}
}{% if KOMA class
  \KOMAoptions{parskip=half}}
\makeatother
\usepackage{longtable,booktabs,array}
\usepackage{calc} % for calculating minipage widths
\usepackage{etoolbox}
\makeatletter
\patchcmd\longtable{\par}{\if@noskipsec\mbox{}\fi\par}{}{}
\makeatother
\IfFileExists{footnotehyper.sty}{\usepackage{footnotehyper}}{\usepackage{footnote}}
\makesavenoteenv{longtable}
\providecommand{\tightlist}{%
  \setlength{\itemsep}{0pt}\setlength{\parskip}{0pt}}
\usepackage[numbers,sort]{natbib}
\usepackage{bookmark}
\IfFileExists{xurl.sty}{\usepackage{xurl}}{} % add URL line breaks if available
\hypersetup{
  pdftitle={Fixed or Deleted? External Feedback Lets Cheaper AI Coding Agents Repair Rule Violations},
  pdfauthor={Laura Wynter  School of Computing and Information Systems, Singapore Management University, Singapore  and CodeAssure Labs, Singapore},
  colorlinks=true,
  linkcolor={Maroon},
  filecolor={Maroon},
  citecolor={Blue},
  urlcolor={Blue},
  pdfcreator={LaTeX via pandoc}}

\title{Fixed or Deleted? External Feedback Lets Cheaper AI Coding Agents
Repair Rule Violations}
\author{Laura Wynter\thanks{Corresponding author: lwynter@smu.edu.sg} \\
\small School of Computing and Information Systems, Singapore Management
University, Singapore \\ \small and CodeAssure Labs, Singapore}
\date{November 2026}

\begin{document}
\maketitle
\begin{abstract}
Many defects violate project rules that the code itself does not reveal,
such as layering constraints or required tests. We study whether LLM
coding agents repair such violations better when an external checker
reports them with their location and a repair goal. Three models of
different sizes, driven by an AI agent, worked on 25 pull requests with
31 planted defects in ten open-source projects, plus 5 correct controls,
five times alone and five times with findings from CodeAssure (CA), a
commercial checker. Counting a run as repaired only when the code also
compiles and keeps its feature, CA findings raised the repair rate from
26\% to 69\% for the small model, from 46\% to 90\% for the medium model
and from 45\% to 93\% for the large model (pull-request-level
p \textless{} 0.001). Agents often cleared a defect by deleting a newly
added file, which metrics based on whether the defect is gone count as a
success; the largest model alone did so in 44 of 125 runs. CA's check
for deleted files largely prevents this: with the check turned off,
removals rose from 7 to 19, 2 to 29 and 0 to 17 of 125 runs for the
small, medium and large models. The CA workflow also outperformed GitHub
Copilot code review (p \textless{} 0.001), and nine free static-analysis
tools together detected 8 of the 31 defects, against all 31 for CA.
With CA, the medium model repaired 90\% of runs, against 45\% for the
largest model alone, at about one fifth of the mean cost per run (\$0.17
against \$0.91).
\end{abstract}

\section{Introduction}\label{introduction}

Coding agents built on large language models (LLMs) now carry out
multi-step software tasks with little supervision. They read a
repository, edit files, run commands and commit. Benchmarks such as
SWE-bench \citep{jimenez2024} measure whether an agent can resolve an
issue so that a project's tests pass. Many requirements on a change,
however, are not covered by tests. Mature projects have architectural
rules (e.g., which layer may depend on which), conventions (e.g., no
console output in service code, or no raw threads in the storage layer)
and process requirements (e.g., a new command must ship with its test).
Such rules are often written only in contributor guides, or not at all,
and their erosion over time is well studied in software architecture
research \citep{desilva2012}.

An agent asked to review its own change must infer these rules from the
code. Prior work suggests that self-review without an external signal is
unreliable. LLMs often fail to correct their own reasoning
\citep{huang2024}, and the benefit of self-repair in code generation
depends on the quality of the feedback \citep{olausson2024}.
Deterministic tools can supply that signal. Architecture conformance
checking has a long history \citep{murphy1995}, and recent work combines
static-analysis findings with LLM repair \citep{wadhwa2024}.

This paper reports a controlled study of coding agents that receive such
feedback. Specifically, we planted defects in pull requests (PRs) on ten widely used
open-source projects written in C++, Rust, Python, TypeScript and Java.
Three  models of different size within an AI coding agent then repaired them, with and
without findings from an external checker. We compare this workflow with
GitHub Copilot code review, and we further compare the checker's detection with that of
free static-analysis tools run out of the box. We ask four questions.

\begin{itemize}
\tightlist
\item
  \textbf{RQ1.} Does external architecture feedback improve the rate at
  which coding agents repair rule violations, and does the effect depend
  on model size?
\item
  \textbf{RQ2.} How do agents fail, and in particular how often do they remove a defect by removing the feature that contains it, and does a deletion check prevent this?
\item
  \textbf{RQ3.} Does  external feedback give cost savings? 
\item
  \textbf{RQ4.} Do widely used free static-analysis tools, run out of
  the box, already detect these defects?
\end{itemize}

In answering the above questions, our contributions are the following.

\begin{itemize}
\tightlist
\item
  Evidence that a workflow with located external feedback, here from one
  commercial checker, raises the repair rate by 43 to 48 percentage points
  when the repaired code must also compile (47 to 58 points under the static scorer alone).
\item
  The measurement of \emph{feature deletion}, in which the agent removes
  a file the pull request added. It is a frequent strategy, and a naive
  success metric counts it as a fix. An ablation shows 
  that CA's deletion check largely prevents it: with the check turned off, 
  removals rise from 7 to 19, 2 to 29 and 0 to 17 of 125 runs for Haiku, Sonnet and Fable,
  and the repair rate falls by 8 to 21 points, 
  a drop significant at the pull-request level for two of the three models.
\item
  Evidence that agents working alone, asked to review a change that
  contains no defect, usually edit its code anyway. In the CA workflow,
  which reported no actionable findings on these changes, they made no
  code edits.
\item
  A comparison on the same pull requests with GitHub Copilot code 
  review in its default configuration.
\item
  A detection comparison with nine free static-analysis tools run out of
  the box (Semgrep, SonarQube Community, and standard linters and
  compilers for each language), which together detected 8 of the 31
  planted defects.
\item
  A cost analysis showing that a mid-size model, given external 
  feedback and the deletion check, repaired twice as many runs as the 
  largest model working alone, at about one fifth of its mean cost per run.
\item
  A build and feature check of 851 scored repairs showing that the gains hold when
  code must compile: with CA findings, the share of runs producing a repair that builds and
  keeps its feature rises from 26\% to 69\% for Haiku, 46\% to 90\% for Sonnet and 45\% to 93\% for Fable. Every Java, TypeScript and Python
  repair compiles, and in the CA workflow no Java or TypeScript repair failed in any way.
\end{itemize}

Section \ref{sec:related} reviews related work and Section
\ref{sec:method} describes the method. Section \ref{sec:results} 
presents the results, including an ablation of CA's deletion check. 
Section \ref{sec:discussion} discusses the main observations from the results. 
Section \ref{sec:arms} compares the CA workflow with
GitHub Copilot code review, and Section \ref{sec:ablation} compares
detection by free static-analysis tools. Section \ref{sec:threats}
discusses threats to validity, and Section \ref{sec:conclusion}
concludes.

\section{Related work}\label{sec:related}

\textbf{Agent benchmarks.} SWE-bench \citep{jimenez2024} and its
human-validated subset \citep{openai2024verified} evaluate issue
resolution by running project tests. Agent scaffolds such as SWE-agent
\citep{yang2024} and Agentless \citep{xia2024} improved results on these
benchmarks. Our setting differs in two ways. The defects are rule
violations that tests usually do not detect, and we also measure
behaviour on changes that contain no defect. Locating the files a change
must touch is a large part of an agent's work, and how an agent explores
the repository affects how well it does so \citep{fattha2026}. Located
findings give the agent this information directly.

\textbf{Self-correction and feedback.} Self-refinement can improve
outputs when the feedback is informative \citep{madaan2023, shinn2023},
but intrinsic self-correction without an external signal often fails or
makes results worse \citep{huang2024}. For code, Olausson et al.
\citep{olausson2024} find that repair is limited by the quality of the
feedback. Aggregating several reasoning traces is another route to
reliability, and can preserve a correct minority view that majority
voting would discard \citep{wynter2026evidence}. Our Alone workflow is a
form of intrinsic self-review. Our +CA workflow adds external, located
feedback that the model would otherwise have to infer.

\textbf{Static analysis with LLM repair.} Wadhwa et al.
\citep{wadhwa2024} pair static-analysis warnings with an LLM proposer
and an LLM ranker to resolve code-quality issues. Blyth et al.
\citep{blyth2025} feed linter and security-scanner warnings back to an
LLM to improve generated code beyond functional correctness, and
CodeCureAgent \citep{joos2025} uses an agent to classify and repair
static-analysis warnings. Su et al. \citep{su2026} find that LLMs detect
violations of explicit architecture decisions that can be inferred from
the code well, and implicit ones less well. Architecture conformance
checking, from reflexion models \citep{murphy1995} to rule libraries run
in continuous integration, detects violations of declared module
dependencies. We study what happens when such findings reach an
autonomous agent through a tool interface and the agent is left to act
on them.

\textbf{Giving agents project knowledge.} Project rules can reach an
agent in several ways. They can be built into the model by fine-tuning
\citep{lee2024knowledge}, which is costly to keep current as rules
change. They can also be written as natural-language instruction or
skill files that the agent reads at run time. Gloaguen et al.
\citep{gloaguen2026} found that repository context files such as
AGENTS.md did not generally improve task success and raised inference
cost by over 20\%, although agents did follow the instructions in them.
Declarative skills of this kind can outperform hand-coded workflows when
the relevant knowledge is retrieved reliably \citep{lim2026skills}. In
this study, the rules are checked outside the agent and violations are
reported to it.

\textbf{Specification gaming.} Optimisers that meet an objective as
literally stated while missing its purpose are well documented
\citep{krakovna2020}. Coding agents have been observed to delete or edit
failing tests, or hard-code expected outputs, instead of fixing the
underlying problem \citep{zhong2025impossible, gabor2025evilgenie}.
Deleting a feature to clear a finding is another example. The checker's
objective (no violations) is met, but the developer's intent (a working
feature without violations) is not. In reinforcement learning,
constraints can be enforced from outside the learned policy rather than
left to it, for example by combining it with stochastic programming
\citep{ghosh2022}. External checking plays a similar role for coding agents, and keeping the feature has to be one of the constraints. CA therefore checks for removed features (Section \ref{sec:design}).

\section{Method}\label{sec:method}

\subsection{Pull requests and planted
defects}\label{pull-requests-and-planted-defects}

We used forks of ten widely used open-source projects. By language, 11
pull requests are in C++, 9 in Rust, 6 in Python, 2 in TypeScript and 2
in Java. Each pull request adds a small, plausible feature written in
the project's style. We planted defects in 25 of them and left 5
correct. Table 1 summarises the defect classes. The five multi-defect
pull requests contain 11 defects in total, in separate files, so that a
single edit cannot fix two of them by accident. Twenty pull requests
come from an earlier single-run study by the same author, and ten were
written for this study.

\begin{longtable}[]{@{}
  >{\raggedright\arraybackslash}p{(\linewidth - 6\tabcolsep) * \real{0.2000}}
  >{\raggedright\arraybackslash}p{(\linewidth - 6\tabcolsep) * \real{0.3200}}
  >{\raggedright\arraybackslash}p{(\linewidth - 6\tabcolsep) * \real{0.4100}}
  >{\raggedright\arraybackslash}p{(\linewidth - 6\tabcolsep) * \real{0.0700}}@{}}
\caption{Pull requests by defect class.}\tabularnewline
\toprule\noalign{}
\begin{minipage}[b]{\linewidth}\raggedright
Class
\end{minipage} & \begin{minipage}[b]{\linewidth}\raggedright
Description
\end{minipage} & \begin{minipage}[b]{\linewidth}\raggedright
Example
\end{minipage} & \begin{minipage}[b]{\linewidth}\raggedright
PRs
\end{minipage} \\
\midrule\noalign{}
\endfirsthead
\toprule\noalign{}
\begin{minipage}[b]{\linewidth}\raggedright
Class
\end{minipage} & \begin{minipage}[b]{\linewidth}\raggedright
Description
\end{minipage} & \begin{minipage}[b]{\linewidth}\raggedright
Example
\end{minipage} & \begin{minipage}[b]{\linewidth}\raggedright
PRs
\end{minipage} \\
\midrule\noalign{}
\endhead
\bottomrule\noalign{}
\endlastfoot
Code-visible & Detectable from the code alone & A renamed function with
callers left behind & 3 \\
Convention & A project-specific rule about what code may do where &
Console output in service code, raw threads in the storage layer & 8 \\
Architecture rule & A declared layering or process constraint & A
data-layer module importing a web route, a command without its required
test & 9 \\
Multiple & Two or three defects from different classes & & 5 \\
Clean & No defect & & 5 \\
\end{longtable}

For each defect we wrote a scorer specification before opening the pull
request. Its defect predicate tests only whether the planted defect is
present, and we checked that it finds the defect on the pull request's
head and not on its base. The run outcomes in Section \ref{sec:outcomes}
add rules about files the pull request added, which do not exist on the
base.

\subsection{Agent and models}\label{agent-and-models}

All runs used Claude Code (version 2.1.285) as the AI coding agent in non-interactive mode,
with no human in the loop. The agent could read and edit files and use a
shell. No build toolchain was installed, so agents could not compile
code or run tests, and fixes were judged by inspecting the final commit
(Section \ref{sec:outcomes}). The project's rule files were read-only
for the agent. We tested three models of different size and price,
the smallest model is Claude Haiku 4.5, Claude Sonnet 5.5 is the medium model 
and the largest model is Claude Fable 5.1 (API
identifiers \texttt{claude-haiku-4-5-20251001},
\texttt{claude-sonnet-5-5} and \texttt{claude-fable-5-1}).

\subsection{Experiment design}\label{sec:design}

The main experiment compares two workflows, \emph{Alone} and \emph{+CA},
which differ in what the agent is asked to do and what information it
receives. To measure the contribution of CA's deletion check, we also ran +CA with that check turned off, as an ablation (Section \ref{sec:deletion}). Section \ref{sec:arms} adds a comparison with GitHub
Copilot code review, and Section \ref{sec:ablation} compares detection
alone with free static-analysis tools.

\textbf{Alone.} The agent received a clone of the repository on the pull
request branch, with an instruction to review the pull request's changes
against the project's conventions and structure, fix any problems the
pull request introduces, keep the change minimal and idiomatic, and
commit and push. Before these runs we removed every trace of the checker
from the repository and its history. Four of the ten projects ship agent
instruction files written by their maintainers (AGENTS.md or CLAUDE.md),
which we left in place in both workflows. Some of these files describe
rules that our defects violate, in one case explicitly (a ban on
\texttt{unsafe} code in one directory). In those projects, an agent
working alone could read part of the relevant rules.

\textbf{+CA.} The agent was instructed to fix the findings that
CodeAssure (CA) reported on the pull request, which it retrieved through
the Model Context Protocol (MCP). Each finding identified where a
project-rule violation occurred and stated the repair goal, without
supplying a patch. After its first attempt, the agent could take one
more repair pass if actionable findings remained. CA is a commercial
checker developed by the author's company (Section \ref{sec:threats}).
Besides AI-written code, it checks staff-built agents and automations
and AI assistants connected to an organisation's systems, against
security risks and the organisation's own rules. This study uses only
its checks of AI-written code. Because agents are known to satisfy checks by deleting or editing the code under test (Section \ref{sec:related}), CA also gives the agent feedback when a file the pull request added has been removed.
Nothing in the design is specific to CA, and any checker that reports
located, rule-level findings to an agent could be evaluated the same
way.

Each run worked on its own copy of the pull request, on a new branch
whose head was an empty commit on top of the original head, so every run
started from an identical tree. Each of the 30 pull requests was run
five times with each of the nine combinations of model and configuration (Alone, +CA, and +CA with the deletion check turned off), giving 1,350 runs, 1,125 on defective pull requests and 225 on clean ones.
Runs of the same configuration on the same pull request were scheduled
back to back.

\subsection{Outcome measures}\label{sec:outcomes}

\textbf{Scoring.} After the agent's last commit, a scorer applies the
specification for each planted defect. The study team implemented the
scorer separately from the checker, and it does not use the checker. It
gives one of three outcomes.

\begin{itemize}
\tightlist
\item
  \emph{Repaired.} Every defect predicate passes, and every non-test
  file the pull request added is still present and still contains code.
\item
  \emph{Feature removed.} Every defect predicate passes, but at least
  one non-test file the pull request added is gone or has been emptied
  of code, whether or not it carried the defect. A file counts as
  emptied when nothing is left but blank lines, comments, braces and
  import, include or namespace lines. A file does not count as gone if
  git reports it as renamed (at least 50\% similarity) or a file with
  the same name exists elsewhere in the tree. For the two pull requests
  whose feature is a rename, reverting the rename also counts as feature
  removed. The rules for emptied files and reverted renames were added
  after a later review of scored outcomes, and all runs were rescored
  with them. Nine runs changed from repaired to feature removed.
  ``Feature removed'' is our shorthand for the removal or emptying of a
  newly added file, or the reverting of a rename. A surviving file that
  still contains code does not show that the feature still works
  (Section \ref{sec:threats}).
\item
  \emph{Not repaired.} At least one defect predicate fails, whether or
  not a file was also deleted. A multi-defect run with one defect still
  open and a file deleted is therefore not repaired.
\end{itemize}

A multi-defect pull request counts as repaired only if all its defects
are repaired, so all rates in this paper are shares of runs, not of the
31 individual defects. We report two measures. The \emph{repair rate},
our primary measure, is the share of runs repaired. The
\emph{defect-gone rate} is the share of runs that are repaired or
feature removed, that is, in which the defect is gone by any route. The
two differ by the runs in which the agent removed the feature, which is
not an acceptable repair. The scorer does not build the code, so after
scoring we also built every repaired run and checked that it keeps its
feature (Section \ref{sec:buildcheck}). The \emph{build-checked repair
rate}, the share of runs repaired and passing that check, is our headline
measure. Where the text gives a repair rate without qualification, it is
the static scorer's.

On clean pull requests we classify the final commit as left alone (no
change, or changes to tests or comments only), a code edit, or a feature
removal.

\textbf{Cost.} The cost of a run is the agent's cost as reported by
Claude Code, over both sessions where there were two, plus the checker's
usage in CA workflows. Infrastructure, licences and other expenses are
not included. The \emph{cost per change} is the mean cost over runs on
defective pull requests. Prices are list prices for the models' public
application programming interfaces (APIs) in November 2026.

\textbf{Agreement.} As a check on the scorer, we compare its defect-gone
verdict with the checker's own verdict on the same final commit.

\subsection{Statistical analysis}\label{statistical-analysis}

Rates are reported with counts and 95\% Wilson score intervals
\citep{wilson1927} over the 125 runs of a configuration. The five runs
of one pull request share its difficulty and were scheduled back to
back, so they are not independent, and we treat the pull request as the
unit of inference. We therefore also report 95\% intervals from a
cluster bootstrap that resamples pull requests \citep{efron1993}, which
better reflects the uncertainty for new pull requests. Cost intervals
use the same bootstrap. All bootstraps draw 10,000 resamples with one
fixed seed, so a configuration that appears in two tables has the same
interval in both. When every run of a configuration succeeds, the
cluster-bootstrap interval collapses to a point. This reflects the
sample, not certainty about new pull requests.

Paired comparisons first average the runs of each pull request. We
report the mean difference per pull request with a paired bootstrap
interval, and use an exact two-sided paired permutation (sign-flip) test
over pull requests as the primary test. When few pull requests differ,
this test cannot reach small p-values. With four differing pull
requests, for example, the smallest possible p is 0.125. As a secondary
measure we report an exact McNemar test \citep{mcnemar1947} on runs
matched by pull request and repetition. Because it treats runs as
independent, it overstates significance. 
Agreement between scorer and checker is summarised with Cohen's
kappa \citep{cohen1960}.

\section{Results}\label{sec:results}

\subsection{Repair rates (RQ1)}\label{repair-rates-rq1}

Table 2 gives the main results. With CA findings, the build-checked repair rate rose
for every model: Haiku from 26\% (32 of 125 runs) to 69\% (86), Sonnet from 46\% (57) to 90\% (113), and Fable from 45\% (56) to 93\% (116). Under the static scorer alone, the rates rose from 28\% (35) to 86\% (107), from 47\% (59) to 94\% (118), and from 45\% (56) to 100\% (125). As expected with 25 pull requests, the cluster-bootstrap intervals are wider than the Wilson intervals, but they do not overlap between the two workflows for any model. The paired comparisons, which control for the difficulty of each pull request, are in Table 3.

\begin{longtable}[]{@{}
  >{\raggedright\arraybackslash}p{(\linewidth - 12\tabcolsep) * \real{0.1950}}
  >{\raggedright\arraybackslash}p{(\linewidth - 12\tabcolsep) * \real{0.1300}}
  >{\raggedright\arraybackslash}p{(\linewidth - 12\tabcolsep) * \real{0.0950}}
  >{\raggedright\arraybackslash}p{(\linewidth - 12\tabcolsep) * \real{0.1550}}
  >{\raggedright\arraybackslash}p{(\linewidth - 12\tabcolsep) * \real{0.1550}}
  >{\raggedright\arraybackslash}p{(\linewidth - 12\tabcolsep) * \real{0.0850}}
  >{\raggedright\arraybackslash}p{(\linewidth - 12\tabcolsep) * \real{0.1850}}@{}}
\caption{Outcomes on the 25 defective pull requests, 125 runs per
configuration. Build-checked: runs repaired that also compile and keep
their feature (Section \ref{sec:buildcheck}), with Wilson and
cluster-bootstrap 95\% intervals. Static: repair rate under the static
scorer alone, with Wilson 95\% interval. Removed: runs with the feature
removed. Cost: mean per change with cluster-bootstrap 95\%
interval.}\tabularnewline
\toprule\noalign{}
\begin{minipage}[b]{\linewidth}\raggedright
Configuration
\end{minipage} & \begin{minipage}[b]{\linewidth}\raggedright
Build-checked
\end{minipage} & \begin{minipage}[b]{\linewidth}\raggedright
(cluster)
\end{minipage} & \begin{minipage}[b]{\linewidth}\raggedright
Static
\end{minipage} & \begin{minipage}[b]{\linewidth}\raggedright
Defect-gone rate
\end{minipage} & \begin{minipage}[b]{\linewidth}\raggedright
Removed
\end{minipage} & \begin{minipage}[b]{\linewidth}\raggedright
Cost per change
\end{minipage} \\
\midrule\noalign{}
\endfirsthead
\toprule\noalign{}
\begin{minipage}[b]{\linewidth}\raggedright
Configuration
\end{minipage} & \begin{minipage}[b]{\linewidth}\raggedright
Build-checked
\end{minipage} & \begin{minipage}[b]{\linewidth}\raggedright
(cluster)
\end{minipage} & \begin{minipage}[b]{\linewidth}\raggedright
Static
\end{minipage} & \begin{minipage}[b]{\linewidth}\raggedright
Defect-gone rate
\end{minipage} & \begin{minipage}[b]{\linewidth}\raggedright
Removed
\end{minipage} & \begin{minipage}[b]{\linewidth}\raggedright
Cost per change
\end{minipage} \\
\midrule\noalign{}
\endhead
\bottomrule\noalign{}
\endlastfoot
Haiku 4.5 alone & 26\% {[}19, 34{]} & {[}11, 41{]} & 28\% {[}21, 36{]} & 38\% {[}30, 47{]} &
13 & \$0.18 {[}0.14, 0.23{]} \\
Haiku 4.5 + CA & 69\% {[}60, 76{]} & {[}54, 82{]} & 86\% {[}78, 91{]} & 91\% {[}85, 95{]} & 7 & \$0.42 {[}0.31, 0.53{]} \\
Sonnet 5.5 alone & 46\% {[}37, 54{]} & {[}28, 63{]} & 47\% {[}39, 56{]} & 72\% {[}64, 79{]}
& 31 & \$0.06 {[}0.06, 0.07{]} \\
Sonnet 5.5 + CA & 90\% {[}84, 94{]} & {[}79, 99{]} & 94\% {[}89, 97{]} & 96\% {[}91, 98{]} & 2 & \$0.17 {[}0.14, 0.19{]} \\
Fable 5.1 alone & 45\% {[}36, 54{]} & {[}28, 62{]} & 45\% {[}36, 54{]} & 80\% {[}72, 86{]} &
44 & \$0.91 {[}0.76, 1.08{]} \\
Fable 5.1 + CA & 93\% {[}87, 96{]} & {[}83, 100{]} & 100\% {[}97, 100{]} & 100\% {[}97, 100{]} & 0 & \$1.73 {[}1.32, 2.18{]} \\
\end{longtable}

Table 3 gives the paired comparisons under the static scorer; Section \ref{sec:buildcheck} repeats them for the build-checked rate, where CA findings raised the repair rate by 43 to 48 points for every model (p \textless{} 0.001). Under the static scorer, for each model CA findings raised the repair rate (pull-request-level p \textless{} 0.001) and the defect-gone rate (p \(\le\) 0.016). Sonnet with CA findings also had a higher repair rate than Fable working alone (p \textless{} 0.001), but not a significantly higher defect-gone rate (p = 0.062). The same holds for Haiku with CA findings against Fable alone (p = 0.001 and p = 0.27). When repairs must also compile, Haiku with CA findings remains ahead of Fable alone, by a margin that is no longer significant (Section \ref{sec:buildcheck}).

\begin{longtable}[]{@{}
  >{\raggedright\arraybackslash}p{(\linewidth - 10\tabcolsep) * \real{0.3400}}
  >{\raggedright\arraybackslash}p{(\linewidth - 10\tabcolsep) * \real{0.2000}}
  >{\raggedright\arraybackslash}p{(\linewidth - 10\tabcolsep) * \real{0.1200}}
  >{\raggedright\arraybackslash}p{(\linewidth - 10\tabcolsep) * \real{0.0800}}
  >{\raggedright\arraybackslash}p{(\linewidth - 10\tabcolsep) * \real{0.1600}}
  >{\raggedright\arraybackslash}p{(\linewidth - 10\tabcolsep) * \real{0.1000}}@{}}
\caption{Paired comparisons over the 25 defective pull requests. Mean
difference: mean per-pull-request difference with paired bootstrap 95\%
interval. ``A only / B only'' counts runs, matched by pull request and
repetition, that A repaired and B did not (for the defect-gone rate, in
which the defect was gone for A and not for B). p: exact two-sided
paired permutation test over pull requests (primary). The secondary
run-level McNemar test gives p \textless{} 0.001 for every comparison
except the defect-gone rate for Haiku + CA against Fable alone (p = 0.029).}\tabularnewline
\toprule\noalign{}
\begin{minipage}[b]{\linewidth}\raggedright
A vs B
\end{minipage} & \begin{minipage}[b]{\linewidth}\raggedright
Repair rate: mean difference {[}95\%{]}
\end{minipage} & \begin{minipage}[b]{\linewidth}\raggedright
A only / B only
\end{minipage} & \begin{minipage}[b]{\linewidth}\raggedright
p
\end{minipage} & \begin{minipage}[b]{\linewidth}\raggedright
Defect-gone rate: A only / B only
\end{minipage} & \begin{minipage}[b]{\linewidth}\raggedright
p
\end{minipage} \\
\midrule\noalign{}
\endfirsthead
\toprule\noalign{}
\begin{minipage}[b]{\linewidth}\raggedright
A vs B
\end{minipage} & \begin{minipage}[b]{\linewidth}\raggedright
Repair rate: mean difference {[}95\%{]}
\end{minipage} & \begin{minipage}[b]{\linewidth}\raggedright
A only / B only
\end{minipage} & \begin{minipage}[b]{\linewidth}\raggedright
p
\end{minipage} & \begin{minipage}[b]{\linewidth}\raggedright
Defect-gone rate: A only / B only
\end{minipage} & \begin{minipage}[b]{\linewidth}\raggedright
p
\end{minipage} \\
\midrule\noalign{}
\endhead
\bottomrule\noalign{}
\endlastfoot
Haiku + CA vs Haiku alone & +58 pts {[}38, 74{]} & 77 / 5 & \textless{} 0.001 & 72 / 6 & \textless{} 0.001 \\
Sonnet + CA vs Sonnet alone & +47 pts {[}28, 66{]} & 62 / 3 & \textless{} 0.001 & 33 / 3 & 0.014 \\
Fable + CA vs Fable alone & +55 pts {[}38, 72{]} & 69 / 0 & \textless{} 0.001 & 25 / 0 & 0.016 \\
Sonnet + CA vs Fable alone & +50 pts {[}33, 66{]} & 62 / 0 & \textless{} 0.001 & 25 / 5 & 0.062 \\
Haiku + CA vs Fable alone & +41 pts {[}20, 60{]} & 59 / 8 & 0.001 & 25 / 11 & 0.27 \\
\end{longtable}

\textbf{By defect class.} Table 4 shows that the effect is concentrated
where the code does not reveal the rule. Working alone, the models
repaired 14 or 15 of the 15 code-visible runs. On architecture rules, the models alone repaired between 5 (Haiku) and 19 (Sonnet) of 45 runs, and between 35 and 45 with CA findings. With CA findings, Sonnet and Fable repaired all 40 convention runs and Haiku 38. On multi-defect pull requests, the models alone repaired every defect in 4 to 8 of 25 runs, and with CA findings in 22 to 25.

\begin{longtable}[]{@{}
  >{\raggedright\arraybackslash}p{(\linewidth - 8\tabcolsep) * \real{0.3200}}
  >{\raggedright\arraybackslash}p{(\linewidth - 8\tabcolsep) * \real{0.1700}}
  >{\raggedright\arraybackslash}p{(\linewidth - 8\tabcolsep) * \real{0.1700}}
  >{\raggedright\arraybackslash}p{(\linewidth - 8\tabcolsep) * \real{0.1700}}
  >{\raggedright\arraybackslash}p{(\linewidth - 8\tabcolsep) * \real{0.1700}}@{}}
\caption{Runs repaired, by defect class.}\tabularnewline
\toprule\noalign{}
\begin{minipage}[b]{\linewidth}\raggedright
Configuration
\end{minipage} & \begin{minipage}[b]{\linewidth}\raggedright
Code-visible
\end{minipage} & \begin{minipage}[b]{\linewidth}\raggedright
Convention
\end{minipage} & \begin{minipage}[b]{\linewidth}\raggedright
Architecture rule
\end{minipage} & \begin{minipage}[b]{\linewidth}\raggedright
Multiple
\end{minipage} \\
\midrule\noalign{}
\endfirsthead
\toprule\noalign{}
\begin{minipage}[b]{\linewidth}\raggedright
Configuration
\end{minipage} & \begin{minipage}[b]{\linewidth}\raggedright
Code-visible
\end{minipage} & \begin{minipage}[b]{\linewidth}\raggedright
Convention
\end{minipage} & \begin{minipage}[b]{\linewidth}\raggedright
Architecture rule
\end{minipage} & \begin{minipage}[b]{\linewidth}\raggedright
Multiple
\end{minipage} \\
\midrule\noalign{}
\endhead
\bottomrule\noalign{}
\endlastfoot
Haiku 4.5 alone & 15/15 & 11/40 & 5/45 & 4/25 \\
Sonnet 5.5 alone & 15/15 & 20/40 & 19/45 & 5/25 \\
Fable 5.1 alone & 14/15 & 21/40 & 13/45 & 8/25 \\
Haiku 4.5 + CA & 11/15 & 38/40 & 35/45 & 23/25 \\
Sonnet 5.5 + CA & 15/15 & 40/40 & 41/45 & 22/25 \\
Fable 5.1 + CA & 15/15 & 40/40 & 45/45 & 25/25 \\
\end{longtable}

Working alone, the models do not rank by size. Sonnet repaired more
architecture-rule runs than Fable (19 against 13 of 45), mostly because
of deletion. Fable removed the feature in 24 of its 45 architecture-rule
runs. On the defect-gone rate, the two are close (36 and 37 of 45).

On code-visible defects, Haiku with CA findings repaired 11 of 15 runs,
against 15 of 15 alone. All four failures are on one pull request, on
which the checker reported the defect gone and the scorer did not.
Section \ref{sec:checker-failures} explains why.

\subsection{Feature deletion and the deletion-check ablation (RQ2)}\label{sec:deletion}

Removing the feature was common when agents worked alone (Table 2, column
``Removed''). Fable did so in 44 of 125 runs (35\%), Sonnet in 31 (25\%)
and Haiku in 13. Deletions occurred mostly on architecture-rule and
multi-defect pull requests, where the planted defect usually sat in a file
the pull request added. When the code that carries the defect is new,
deleting it satisfies the instruction ``fix any problems this pull request
introduces'' in a literal sense, and is the shortest route to a clean
review. With CA, which reports removed features, removals fell to 7 for
Haiku, 2 for Sonnet and 0 for Fable.

\textbf{Ablation of the deletion check.} To measure what the deletion
check contributes, we ran the +CA workflow with that check turned off,
with the same 30 pull requests, five runs per model, and the same agent,
instructions, session limit and scorer (Table 5). Without the check,
CA's other findings barely reduced removal for Sonnet (29 runs, against 31
working alone) and increased it for Haiku (from 13 to 19), and Fable removed the feature in
17 runs, against 0 with the check.

\begin{longtable}[]{@{}
  >{\raggedright\arraybackslash}p{(\linewidth - 12\tabcolsep) * \real{0.2400}}
  >{\raggedright\arraybackslash}p{(\linewidth - 12\tabcolsep) * \real{0.1400}}
  >{\raggedright\arraybackslash}p{(\linewidth - 12\tabcolsep) * \real{0.1100}}
  >{\raggedright\arraybackslash}p{(\linewidth - 12\tabcolsep) * \real{0.1500}}
  >{\raggedright\arraybackslash}p{(\linewidth - 12\tabcolsep) * \real{0.0800}}
  >{\raggedright\arraybackslash}p{(\linewidth - 12\tabcolsep) * \real{0.0800}}
  >{\raggedright\arraybackslash}p{(\linewidth - 12\tabcolsep) * \real{0.2000}}@{}}
\caption{Ablation of the deletion check: +CA with the check (as in Table 2) and with it turned off, on the 25 defective pull requests (125 runs per configuration). Repair rate: Wilson 95\%
interval, then cluster-bootstrap 95\% interval. Cost: mean per change,
including the checker, with cluster-bootstrap 95\%
interval.}\tabularnewline
\toprule\noalign{}
\begin{minipage}[b]{\linewidth}\raggedright
Configuration
\end{minipage} & \begin{minipage}[b]{\linewidth}\raggedright
Repair rate
\end{minipage} & \begin{minipage}[b]{\linewidth}\raggedright
(cluster)
\end{minipage} & \begin{minipage}[b]{\linewidth}\raggedright
Defect-gone rate
\end{minipage} & \begin{minipage}[b]{\linewidth}\raggedright
Removed
\end{minipage} & \begin{minipage}[b]{\linewidth}\raggedright
2nd session
\end{minipage} & \begin{minipage}[b]{\linewidth}\raggedright
Cost per change
\end{minipage} \\
\midrule\noalign{}
\endfirsthead
\toprule\noalign{}
\begin{minipage}[b]{\linewidth}\raggedright
Configuration
\end{minipage} & \begin{minipage}[b]{\linewidth}\raggedright
Repair rate
\end{minipage} & \begin{minipage}[b]{\linewidth}\raggedright
(cluster)
\end{minipage} & \begin{minipage}[b]{\linewidth}\raggedright
Defect-gone rate
\end{minipage} & \begin{minipage}[b]{\linewidth}\raggedright
Removed
\end{minipage} & \begin{minipage}[b]{\linewidth}\raggedright
2nd session
\end{minipage} & \begin{minipage}[b]{\linewidth}\raggedright
Cost per change
\end{minipage} \\
\midrule\noalign{}
\endhead
\bottomrule\noalign{}
\endlastfoot
Haiku 4.5 + CA, check off & 71\% {[}63, 78{]} & {[}57, 84{]} & 86\% {[}79, 91{]} &
19 & 33 & \$0.37 {[}0.28, 0.46{]} \\
Haiku 4.5 + CA & 86\% {[}78, 91{]} & {[}75, 94{]} & 91\%
{[}85, 95{]} & 7 & 41 & \$0.42 {[}0.31, 0.53{]} \\
Sonnet 5.5 + CA, check off & 74\% {[}65, 81{]} & {[}58, 87{]} & 97\% {[}92, 99{]} &
29 & 19 & \$0.14 {[}0.12, 0.15{]} \\
Sonnet 5.5 + CA & 94\% {[}89, 97{]} & {[}86, 100{]} &
96\% {[}91, 98{]} & 2 & 46 & \$0.17 {[}0.14, 0.19{]} \\
Fable 5.1 + CA, check off & 86\% {[}79, 91{]} & {[}73, 98{]} & 100\% {[}97, 100{]}
& 17 & 7 & \$1.22 {[}1.01, 1.45{]} \\
Fable 5.1 + CA & 100\% {[}97, 100{]} & {[}100, 100{]} & 100\% {[}97, 100{]} & 0 & 18 & \$1.73
{[}1.32, 2.18{]} \\
\end{longtable}

Turning the check off lowered the repair rate for every model, by 14
percentage points for Haiku (95\% interval {[}3, 27{]}, p = 0.036), 21
points for Sonnet ({[}7, 35{]}, p = 0.010) and 14 points for Fable
({[}2, 27{]}, p = 0.12), in paired comparisons with full CA. The paired
intervals exclude zero for all three models, and the pull-request-level
test is significant for Haiku and Sonnet. For Fable the difference comes from
a few pull requests, which limits how small a p-value the test can reach.
The defect-gone rate changed little and not significantly for any model,
so the check turns removals into repairs rather than into unrepaired
defects. It also has a cost: more runs needed a second session (for
Fable, 18 instead of 7), and the cost per change was 13\% to 42\% higher
with the check than without it. The check does what it is designed to do:
agents keep the feature and rewrite the code instead of deleting it. Under
the build check, the repair rate with the check off is 54\% for Haiku, 70\% for
Sonnet and 85\% for Fable, and the check's effect remains significant for Haiku
(15 points, p = 0.023) and Sonnet (21 points, p = 0.010). For
Fable it falls to 8 points (p = 0.25), because some of the rewrites in one
project do not compile (Section \ref{sec:buildcheck}).

Of the 9 runs with CA that still ended with the feature removed, the
checker reported the deletion in 6, but the agent did not restore the
feature in its remaining session. The other 3 were Haiku runs. In two of
them the checker did not report the deletion, and we are investigating
why. In the third, the agent emptied the file instead of deleting it,
which the deletion check does not cover. The 16 runs with CA in which the
defect remained fall into three groups. In ten, the agent did not add a
test that the project requires for a new command or component. Four are
the Haiku runs on the pull request described in Section
\ref{sec:checker-failures}, where the checker reported only one of four
broken calls. In the other two, both Haiku runs on Rust pull requests, a
forbidden dependency between packages was left in place.

An earlier single-run study by the same author reported repair rates of
91\% to 100\% with the same checker, on a different set of 33 changes
and with an earlier Sonnet model. In that study, deleting a file
required an interactive approval that an unattended run could not give,
so agents had to repair the code. This may explain part of the
difference between the two studies, but we did not test it on the pull
requests they share.

\subsection{Clean pull requests}\label{clean-pull-requests}

Table 6 shows behaviour on the five clean pull requests. Asked to review
and fix a correct change, agents working alone usually edited code, in
64\% of runs for Haiku, 100\% for Sonnet and 80\% for Fable, changing 8
to 22 lines on average. Fable also deleted the feature in 5 runs. The checker reported nothing actionable in any of the 75 +CA runs on clean pull requests, and in every one the agent left the change alone, as it also did with the deletion check turned off. These
runs were also much cheaper, because the agent had nothing to do.

\begin{longtable}[]{@{}
  >{\raggedright\arraybackslash}p{(\linewidth - 10\tabcolsep) * \real{0.3000}}
  >{\raggedright\arraybackslash}p{(\linewidth - 10\tabcolsep) * \real{0.1100}}
  >{\raggedright\arraybackslash}p{(\linewidth - 10\tabcolsep) * \real{0.1500}}
  >{\raggedright\arraybackslash}p{(\linewidth - 10\tabcolsep) * \real{0.1100}}
  >{\raggedright\arraybackslash}p{(\linewidth - 10\tabcolsep) * \real{0.1300}}
  >{\raggedright\arraybackslash}p{(\linewidth - 10\tabcolsep) * \real{0.2000}}@{}}
\caption{Outcomes on the 5 clean pull requests, 25 runs per
configuration.}\tabularnewline
\toprule\noalign{}
\begin{minipage}[b]{\linewidth}\raggedright
Configuration
\end{minipage} & \begin{minipage}[b]{\linewidth}\raggedright
Left alone
\end{minipage} & \begin{minipage}[b]{\linewidth}\raggedright
Unnecessary code edit
\end{minipage} & \begin{minipage}[b]{\linewidth}\raggedright
Feature removed
\end{minipage} & \begin{minipage}[b]{\linewidth}\raggedright
Mean lines changed
\end{minipage} & \begin{minipage}[b]{\linewidth}\raggedright
Cost per change
\end{minipage} \\
\midrule\noalign{}
\endfirsthead
\toprule\noalign{}
\begin{minipage}[b]{\linewidth}\raggedright
Configuration
\end{minipage} & \begin{minipage}[b]{\linewidth}\raggedright
Left alone
\end{minipage} & \begin{minipage}[b]{\linewidth}\raggedright
Unnecessary code edit
\end{minipage} & \begin{minipage}[b]{\linewidth}\raggedright
Feature removed
\end{minipage} & \begin{minipage}[b]{\linewidth}\raggedright
Mean lines changed
\end{minipage} & \begin{minipage}[b]{\linewidth}\raggedright
Cost per change
\end{minipage} \\
\midrule\noalign{}
\endhead
\bottomrule\noalign{}
\endlastfoot
Haiku 4.5 alone & 9 & 16 (64\%) & 0 & 7.8 & \$0.11 \\
Sonnet 5.5 alone & 0 & 25 (100\%) & 0 & 8.5 & \$0.06 \\
Fable 5.1 alone & 0 & 20 (80\%) & 5 & 21.6 & \$0.73 \\
Each model + CA & 25 & 0 & 0 & 0.0 & \$0.03 to \$0.27 \\
\end{longtable}

\subsection{Scorer validity}\label{sec:scorer-validity}

Of the 750 runs on defective pull requests of the Alone workflow and of the ablation (Section \ref{sec:deletion}), 748 have both a scorer verdict and a checker verdict, with the checker run
on each final commit after the study. The checker's verdict for one rule
type was not collected in two runs working alone. The scorer and the
checker agreed in 94.8\% of runs (Cohen's kappa 0.84). In 24 runs the
checker reported the defect gone and the scorer did not. Fourteen of
these come from gaps in two of the checker's rules, found during this
study. The other ten are Haiku runs: four on the pull request discussed
in Section \ref{sec:checker-failures}, and six on two Rust pull requests
that we are still investigating. In 15 runs the scorer reported the
defect gone and the checker did not, 4 of them because of an error in
one of the checker's rules. Where the two disagree, we report the
scorer's verdict. Agreement shows that two tools built by the same team
are consistent, not that either is correct. A blinded audit of final
diffs would be a stronger check, and is left for future work.

\subsection{Cost and effort (RQ3)}\label{cost-and-effort-rq3}

Table 7 summarises effort and cost. Including the checker, the cost per change rose from \$0.18 to \$0.42 for Haiku, from \$0.06 to \$0.17 for Sonnet and from \$0.91 to \$1.73 for Fable. Agent steps rose from 43 to 60 for Haiku, from 12 to 16 for Sonnet and from 22 to 34 for Fable. A second session was needed in 33\% of Haiku runs, 37\% of Sonnet runs and 14\% of Fable runs. Of the added cost, 13\% to 42\% comes from the deletion check (Section \ref{sec:deletion}).

\begin{longtable}[]{@{}
  >{\raggedright\arraybackslash}p{(\linewidth - 6\tabcolsep) * \real{0.1905}}
  >{\raggedright\arraybackslash}p{(\linewidth - 6\tabcolsep) * \real{0.2619}}
  >{\raggedright\arraybackslash}p{(\linewidth - 6\tabcolsep) * \real{0.2381}}
  >{\raggedright\arraybackslash}p{(\linewidth - 6\tabcolsep) * \real{0.3095}}@{}}
\caption{Effort and cost per change on defective pull
requests.}\tabularnewline
\toprule\noalign{}
\begin{minipage}[b]{\linewidth}\raggedright
Model
\end{minipage} & \begin{minipage}[b]{\linewidth}\raggedright
Steps, alone / with
\end{minipage} & \begin{minipage}[b]{\linewidth}\raggedright
Runs with a second session
\end{minipage} & \begin{minipage}[b]{\linewidth}\raggedright
Cost per change, alone / with
\end{minipage} \\
\midrule\noalign{}
\endfirsthead
\toprule\noalign{}
\begin{minipage}[b]{\linewidth}\raggedright
Model
\end{minipage} & \begin{minipage}[b]{\linewidth}\raggedright
Steps, alone / with
\end{minipage} & \begin{minipage}[b]{\linewidth}\raggedright
Runs with a second session
\end{minipage} & \begin{minipage}[b]{\linewidth}\raggedright
Cost per change, alone / with
\end{minipage} \\
\midrule\noalign{}
\endhead
\bottomrule\noalign{}
\endlastfoot
Haiku 4.5 & 43 / 60 & 41 of 125 & \$0.18 / \$0.42 \\
Sonnet 5.5 & 12 / 16 & 46 of 125 & \$0.06 / \$0.17 \\
Fable 5.1 & 22 / 34 & 18 of 125 & \$0.91 / \$1.73 \\
\end{longtable}

CA findings change the ranking by cost-effectiveness. Counting only repairs that also compile, Sonnet with CA findings repaired 90\% of runs at a mean of \$0.17 per change, while Fable alone repaired 45\% at \$0.91 (paired difference 46 points, p \textless{} 0.001). The best absolute result, Fable with CA findings at 93\%, cost \$1.73 per change.

\subsection{Build and feature check}\label{sec:buildcheck}

The scorer reads the final commit without building it (Section \ref{sec:outcomes}). To see how many repairs would also
compile, after all runs had been scored we built the final commit of every run scored as repaired on a defective pull
request, in the main study and in the Copilot comparison (851 runs). Rust projects were checked with
\texttt{cargo check}, C++ projects were built with CMake, Java with \texttt{mvn test-compile} and TypeScript with
\texttt{tsc}; Python files were compiled and checked for undefined names with Ruff. The navigation2 pull request needs a
ROS~2 installation, so for its 26 repaired runs we only checked that every project header and symbol the code uses
exists. To separate the agent's changes from the pull request's own state, we also built the base commit and the pull
request's head, and counted only errors absent from the head. Four pull-request heads do not compile on their own, for
reasons unrelated to the planted defect, and in one project the file added by the pull request is not part of the
build; errors inherited from these heads were not counted. We also inspected the diff of every repaired run for stubs
and for features removed in place, which the file-level rule of Section \ref{sec:outcomes} does not detect.

The CA workflow's advantage holds when repairs must also compile. With CA findings, the share of runs that produced a
repair passing the build and feature check rose from 26\% to 69\% for Haiku, from 46\% to 90\% for Sonnet and from
45\% to 93\% for Fable (Table~\ref{tab:buildcheck}). Every Java, TypeScript and Python repair compiled,
in every workflow, and in the CA workflow no Java or TypeScript repair failed the check in any way (0 of 15 runs each;
Table~\ref{tab:buildlang}).

Of the 851 repairs, 75 did not pass: 50 that do not compile, all in Rust and C++; 21 Python repairs that removed the
behaviour the pull request added while keeping its file (a validator deleted from a schema class, with nothing applying
it elsewhere); and 4 stubs in Rust and C++. The compile failures arise where agents worked without a compiler: in Rust,
the language with the highest share (14\% of its repairs), they called functions that do not exist or broke its
thread-safety rules. The in-place removals occurred in two Python pull requests whose planted defect is an import from
the service layer into the schema layer. Deleting the validator satisfies the layering rule, and the deletion check,
which looks for removed or emptied files, does not cover it. Most failures came from the smallest model: Haiku with CA
findings accounts for 21 of the 35 failures in the CA workflow. Table~\ref{tab:buildcheck} compares the two measures by
configuration.

\begin{longtable}[]{@{}lrrrr@{}}
\caption{Build and feature check by language: repairs scored by the static scorer and those that failed the check,
over all workflows, and for +CA with the deletion check only.}\label{tab:buildlang}\tabularnewline
\toprule\noalign{}
Language & Checked & Do not compile & Removed in place or stub & Failed with +CA \\
\midrule\noalign{}
\endfirsthead
\toprule\noalign{}
Language & Checked & Do not compile & Removed in place or stub & Failed with +CA \\
\midrule\noalign{}
\endhead
\bottomrule\noalign{}
\endlastfoot
Java       & 51  & 0  & 0  & 0 of 15 \\
TypeScript & 30  & 0  & 0  & 0 of 15 \\
Python     & 202 & 0  & 21 & 8 of 70 \\
C++        & 305 & 16 & 2  & 8 of 137 \\
Rust       & 263 & 34 & 2  & 19 of 113 \\
\midrule\noalign{}
Total      & 851 & 50 & 25 & 35 of 350 \\
\end{longtable}

Under the stricter measure, all the main results hold. CA findings raise the repair rate for every model (Haiku +43
points [23, 62], Sonnet +45 [26, 64], Fable +48 [31, 65]; pull-request-level p \textless{} 0.001). Sonnet with CA
findings still repairs twice as many runs as Fable alone (90\% against 45\%, p \textless{} 0.001), at about one fifth
of the cost. The CA workflow still outperforms Copilot code review for both models (p $\le$ 0.002), and the deletion
check still raises the repair rate significantly for Haiku (p = 0.023) and Sonnet (p = 0.010). Two secondary
comparisons are less clear. Haiku with CA findings remains ahead of Fable alone (+24 points [$-1$, 47]), but the
difference is no longer significant (p = 0.075). For Fable, the deletion check led the agent to keep files instead of
deleting them, but 9 of the rewrites in one project do not compile, so its effect falls to 8 points ([$-2$, 19],
p = 0.25).

\begin{longtable}[]{@{}llrr@{}}
\caption{Repair rate on the 25 defective pull requests under the static scorer and after the
build and feature check (the headline measure). 125 runs per configuration, 72 for Copilot code review.}\label{tab:buildcheck}\tabularnewline
\toprule\noalign{}
Model & Workflow & Static scorer & Build-checked \\
\midrule\noalign{}
\endfirsthead
\toprule\noalign{}
Model & Workflow & Static scorer & Build-checked \\
\midrule\noalign{}
\endhead
\bottomrule\noalign{}
\endlastfoot
Haiku 4.5  & Alone                      & 35 (28\%)  & 32 (26\%) \\
           & +CA                        & 107 (86\%) & 86 (69\%) \\
           & +CA, deletion check off    & 89 (71\%)  & 67 (54\%) \\
           & Copilot review             & 29 (40\%)  & 23 (32\%) \\
Sonnet 5.5 & Alone                      & 59 (47\%)  & 57 (46\%) \\
           & +CA                        & 118 (94\%) & 113 (90\%) \\
           & +CA, deletion check off    & 92 (74\%)  & 87 (70\%) \\
           & Copilot review             & 33 (46\%)  & 33 (46\%) \\
Fable 5.1  & Alone                      & 56 (45\%)  & 56 (45\%) \\
           & +CA                        & 125 (100\%) & 116 (93\%) \\
           & +CA, deletion check off    & 108 (86\%) & 106 (85\%) \\
\end{longtable}

The check establishes that repairs compile and keep their feature; it does not test behaviour. Most planted features
are small and have no observable behaviour to test, so we did not run project tests.

\section{Discussion}\label{sec:discussion}

\paragraph*{Rule discovery and repair}\label{rule-discovery-and-repair}

The largest gains came from rules that exist only in a project's own
conventions and structure. Working alone, an agent must first discover
such a rule and then decide whether the change violates it. CA findings
do both steps for it. What remains is to repair a located violation
towards a stated goal, which all three models did in most runs. This
agrees with the observation that feedback quality bounds self-repair
\citep{olausson2024}, and suggests that on this kind of task, much of
the difference between models lies in discovering the rules rather than
in repairing violations.

\paragraph*{Model size and review
quality}\label{model-size-and-review-quality}

Working alone, the largest model did not outperform the mid-size one on
the repair rate, and it deleted the feature most often. A more capable
agent may also be more willing to take large actions, including removing
code it judges to be at fault, but we did not test this. In our data,
model size did not predict how well an agent reviewed its own change,
while CA findings improved every model. Other work makes smaller models
sufficient by changing the model, through distillation \citep{kundu2024}
or by combining smaller domain experts \citep{lee2024moe}. Here, the
input to the model changed instead.

\paragraph*{Scoring deletion}\label{sec:scoring-deletion}

In the ablation with the deletion check turned off, a metric that asks only whether the defect is gone would have reported 86\% to 100\% success, against static repair rates of 71\% to 86\%. For Fable alone it would have reported 80\%
instead of 45\%, ranking Fable alone above Sonnet alone (72\%), the
reverse of their order on the repair rate. We recommend that evaluations
of agent repair report deletion as a separate outcome, and that feedback
systems treat the removal of a flagged file as a finding in its own
right, as CA does. Without that finding, CA's other findings barely reduced deletion for Sonnet and did not reduce it for Haiku (Section \ref{sec:deletion}).

\paragraph*{Compiled and interpreted languages}\label{sec:languages}

The build check divides the languages in two. In Java, TypeScript and Python, the languages in which most
AI-written code is produced, every repair compiled, and the remaining failures were Python repairs that met a
layering rule by deleting the behaviour in question rather than moving it. In Rust and C++, agents without a compiler,
above all the smallest model, wrote repairs that meet the repair goal but do not build. Two additions to the feedback
loop follow: returning compiler errors to the agent for compiled languages, and extending the deletion check from
removed files to removed functions and validators.

\paragraph*{Changes with no defect}\label{changes-with-no-defect}

On clean pull requests, agents working alone nearly always changed
something. They appear to read an instruction to ``review and fix any
problems'' as a sign that problems exist. When the checker reported
nothing, the agents stopped, which avoided unnecessary edits and cost
less. This comparison does not separate the effect of the findings from
the difference in instructions. The agent working alone was asked to
review the change broadly, while the +CA agent was asked to fix
findings, of which there were none. We did not test whether findings
prevent unnecessary edits when both agents receive the same review
instruction.

\paragraph*{Failure cases of the checker}\label{sec:checker-failures}

In 24 runs the checker missed a defect that the scorer caught. Section
\ref{sec:scorer-validity} describes the gaps in two of its rules. One
further case shows how a checker's omission can lower an agent's result.

One pull request accounts for Haiku's weaker result on code-visible
defects. It made a new parameter of a shared function required, but left
four existing calls to the function without the new argument. The defect
counts as repaired only when all four calls are corrected. The checker
reported one of the four calls to the agent, and not the other three. In
four of its five +CA runs, Haiku corrected the reported call and
stopped. The checker then reported no remaining findings, so the run
ended with three calls still broken. Working alone, Haiku repaired the
defect in all five runs, either by correcting all four calls or by
giving the new parameter a default value so that the existing calls
stayed valid. Sonnet and Fable, with and without CA findings, gave the
parameter a default value in every run, which fixes all four calls with
one change to the function itself. With CA findings, Haiku repaired only
the call it was shown, treating the findings as a complete list.

An agent that relies on a checker will miss what the checker misses, and
here the smallest model was the one that treated the findings as
complete. In a deployed loop, a missed rule or call site means the agent
stops with a defect still in place. Two changes may reduce this risk.
The checker can report every affected location, including the less
certain ones, and the agent's instructions can ask it to look for
related problems beyond the findings, so that external findings add to
its own review rather than replace it. After the study, the checker was
changed, and it now reports all four calls on this pull request.

\section{Comparison with GitHub Copilot code review}\label{sec:arms}

The main experiment compares external findings with no help at all. A
common alternative is an AI review of the pull request. We therefore ran
a workflow for Haiku and Sonnet, on the same 30 pull requests and
with the same scorer, using GitHub Copilot code review.

\textbf{GitHub Copilot code review (+Copilot).} For each pull request
and repetition, GitHub Copilot reviewed the pull request at its default ``Balanced'' effort, in its default configuration. It saw the same repository as the other workflows, including the maintainers' own instruction files and contributor guides. We did not add custom review instructions (a \texttt{.github/copilot-instructions.md} file), which teams can use to give Copilot their own rules. Its review comments were
given to the agent, which had one session to address them. This reflects
common practice, in which a pull request receives one AI review. To
limit cost, Copilot was run with three repetitions rather than five.

\begin{longtable}[]{@{}
  >{\raggedright\arraybackslash}p{(\linewidth - 10\tabcolsep) * \real{0.1429}}
  >{\raggedright\arraybackslash}p{(\linewidth - 10\tabcolsep) * \real{0.2857}}
  >{\raggedright\arraybackslash}p{(\linewidth - 10\tabcolsep) * \real{0.1607}}
  >{\raggedright\arraybackslash}p{(\linewidth - 10\tabcolsep) * \real{0.0804}}
  >{\raggedright\arraybackslash}p{(\linewidth - 10\tabcolsep) * \real{0.1071}}
  >{\raggedright\arraybackslash}p{(\linewidth - 10\tabcolsep) * \real{0.2232}}@{}}
\caption{Alone, +Copilot and +CA on the 25
defective pull requests, and runs with unnecessary code edits on the 5
clean pull requests. Repair rate: Wilson 95\% interval. The workflows
differ in more than the help given. +Copilot: one session, no custom review instructions, three repetitions, on 24 defective pull requests (72 runs per
model) and the 5 clean pull requests (15 runs per model), one pull
request excluded (see text). +CA: up to two sessions.
The table therefore compares workflows as run.}\tabularnewline
\toprule\noalign{}
\begin{minipage}[b]{\linewidth}\raggedright
Model
\end{minipage} & \begin{minipage}[b]{\linewidth}\raggedright
Workflow
\end{minipage} & \begin{minipage}[b]{\linewidth}\raggedright
Repair rate
\end{minipage} & \begin{minipage}[b]{\linewidth}\raggedright
Removed
\end{minipage} & \begin{minipage}[b]{\linewidth}\raggedright
Clean: code edits
\end{minipage} & \begin{minipage}[b]{\linewidth}\raggedright
Paired vs CA
\end{minipage} \\
\midrule\noalign{}
\endfirsthead
\toprule\noalign{}
\begin{minipage}[b]{\linewidth}\raggedright
Model
\end{minipage} & \begin{minipage}[b]{\linewidth}\raggedright
Workflow
\end{minipage} & \begin{minipage}[b]{\linewidth}\raggedright
Repair rate
\end{minipage} & \begin{minipage}[b]{\linewidth}\raggedright
Removed
\end{minipage} & \begin{minipage}[b]{\linewidth}\raggedright
Clean: code edits
\end{minipage} & \begin{minipage}[b]{\linewidth}\raggedright
Paired vs CA
\end{minipage} \\
\midrule\noalign{}
\endhead
\bottomrule\noalign{}
\endlastfoot
Haiku 4.5 & Alone & 28\% {[}21, 36{]} & 13 & 16 / 25 & 77 / 5, p
\textless{} 0.001 \\
Haiku 4.5 & + Copilot & 40\% {[}30, 52{]} & 13 & 12 / 15 & 38 / 5, p
\textless{} 0.001 \\
Haiku 4.5 & + CA & 86\% {[}78, 91{]} & 7 & 0 / 25 & \\
Sonnet 5.5 & Alone & 47\% {[}39, 56{]} & 31 & 25 / 25 & 62 / 3, p
\textless{} 0.001 \\
Sonnet 5.5 & + Copilot & 46\% {[}35, 57{]} & 9 & 12 / 15 & 37 / 1, p
\textless{} 0.001 \\
Sonnet 5.5 & + CA & 94\% {[}89, 97{]} & 2 & 0 / 25 & \\
\end{longtable}

CA had a higher repair rate than both working alone and Copilot's review, for both models (pull-request-level p
\textless{} 0.001). The paired column gives the runs repaired only with CA against the runs repaired only with the other
workflow, and the pull-request-level p.

Copilot's review did not significantly change the repair rate. In paired
comparisons with the agents working alone, 17 runs were repaired only
with Copilot's review and 9 only without it for Haiku (p = 0.35), and 8
against 12 for Sonnet (p = 0.65). On the clean pull requests, both
models acted on Copilot's comments and edited code in 12 of 15 runs.
Copilot's comments addressed code quality, and in at least one case identified a layering problem from its consequences for the design.

Copilot repeatedly failed to review one pull request. Its full review
did not start, so it fell back to its lightest setting, and later
requests received no review at all. We therefore excluded this pull
request from the Copilot workflow and its paired comparisons. Copilot
runs (repetitions 1 to 3) were paired with repetitions 1 to 3 of the
other workflows on the same pull request. As a sensitivity check,
scoring the excluded pull request as repaired in all three runs, or in
none, gives repair rates of 39\% to 43\% for Haiku and 44\% to 48\% for
Sonnet. Neither bound changes the comparison with working alone (p
\(\ge\) 0.21) or with CA (p \textless{} 0.001).

Copilot's review is general. In its default configuration it reviews against general good practice, while the largest gains from CA findings came on project-specific rules (Section \ref{sec:results}). Giving Copilot such rules through custom review instructions is possible, but those rules would first have to be written for each project, and we did not test it.

\section{Detection by free static-analysis tools}\label{sec:ablation}

We examine whether static analysis tools  would catch these defects
without a dedicated checker. To isolate the role of the CodeAssure checker 
from the AI agent itself, we
compared detection alone without running AI agents. 
We ran nine widely used free tools on each
pull request with their default or standard configuration and no
project-specific rules: Semgrep Community Edition 1.185.0 with its
public \texttt{p/default} and language rule packs; SonarQube Community
Build 26.11 with its default ``Sonar way'' profiles; Pylint 4.1 and Ruff
for Python; ESLint 9 with its recommended TypeScript rules; clang-tidy
23 and cppcheck 2.22 (warning, style, performance and portability
checks) for C++; PMD with its recommended ``quickstart'' rule set for
Java; and, for Rust, a \texttt{cargo\ clippy} build with default lints,
which also runs the compiler. Each tool analysed the files the pull
request changed, plus every file in which a planted defect sits. A tool
detected a defect if it reported a finding in the defect's file within
three lines of the defect, or anywhere in the file for a missing
companion test. We checked every match by hand and rejected three that
were unrelated to the defect: unresolved third-party imports reported by
Pylint, because the projects' dependencies were not installed, and an
\texttt{unsafe} warning that Semgrep raised next to, but not about, a
stale call. For CA we used its first findings on each pull request, as received by the agents at the start of each run.

\begin{longtable}[]{@{}
  >{\raggedright\arraybackslash}p{(\linewidth - 12\tabcolsep) * \real{0.3448}}
  >{\raggedright\arraybackslash}p{(\linewidth - 12\tabcolsep) * \real{0.1954}}
  >{\raggedright\arraybackslash}p{(\linewidth - 12\tabcolsep) * \real{0.0920}}
  >{\raggedright\arraybackslash}p{(\linewidth - 12\tabcolsep) * \real{0.0920}}
  >{\raggedright\arraybackslash}p{(\linewidth - 12\tabcolsep) * \real{0.0920}}
  >{\raggedright\arraybackslash}p{(\linewidth - 12\tabcolsep) * \real{0.0920}}
  >{\raggedright\arraybackslash}p{(\linewidth - 12\tabcolsep) * \real{0.0920}}@{}}
\caption{Detection of the 31 planted defects by tools run out of the
box. In scope: defects in languages the tool analyses. Not covered:
defects in other languages. No tool, including CA, reported any finding
on the lines changed by the 5 clean pull requests.}\tabularnewline
\toprule\noalign{}
\begin{minipage}[b]{\linewidth}\raggedright
Tool
\end{minipage} & \begin{minipage}[b]{\linewidth}\raggedright
Languages
\end{minipage} & \begin{minipage}[b]{\linewidth}\raggedright
In scope
\end{minipage} & \begin{minipage}[b]{\linewidth}\raggedright
Detected
\end{minipage} & \begin{minipage}[b]{\linewidth}\raggedright
Missed in scope
\end{minipage} & \begin{minipage}[b]{\linewidth}\raggedright
Not covered
\end{minipage} & \begin{minipage}[b]{\linewidth}\raggedright
Not caught (of 31)
\end{minipage} \\
\midrule\noalign{}
\endfirsthead
\toprule\noalign{}
\begin{minipage}[b]{\linewidth}\raggedright
Tool
\end{minipage} & \begin{minipage}[b]{\linewidth}\raggedright
Languages
\end{minipage} & \begin{minipage}[b]{\linewidth}\raggedright
In scope
\end{minipage} & \begin{minipage}[b]{\linewidth}\raggedright
Detected
\end{minipage} & \begin{minipage}[b]{\linewidth}\raggedright
Missed in scope
\end{minipage} & \begin{minipage}[b]{\linewidth}\raggedright
Not covered
\end{minipage} & \begin{minipage}[b]{\linewidth}\raggedright
Not caught (of 31)
\end{minipage} \\
\midrule\noalign{}
\endhead
\bottomrule\noalign{}
\endlastfoot
CA & all five & 31 & 31\textsuperscript{b} & 0 & 0 & 0 \\
Rust build and Clippy & Rust & 10 & 6 & 4 & 21 & 25 \\
Semgrep & all five & 31 & 2 & 29 & 0 & 29 \\
Pylint & Python & 7 & 1 & 6 & 24 & 30 \\
SonarQube Community & Python, TypeScript, Java & 9 & 1 & 8 & 22 & 30 \\
Ruff & Python & 7 & 0 & 7 & 24 & 31 \\
ESLint & TypeScript & 1 & 0 & 1 & 30 & 31 \\
clang-tidy & C++ & 12 & 0 & 12 & 19 & 31 \\
cppcheck & C++ & 12 & 0 & 12 & 19 & 31 \\
PMD & Java & 1 & 0 & 1 & 30 & 31 \\
All free tools together & all five & 31 & 8 & 23 & 0 & 23 \\
\end{longtable}

\textsuperscript{b} A defect counts as detected if a tool reported at
least one of its affected lines. Some detections were partial: 7 for CA,
for example 1 of the 4 broken calls in the pull request of Section
\ref{sec:checker-failures}; 3 for the Rust build, which flags the
dependency declaration but not the code that uses it; and 1 for
SonarQube.

Together the free tools detected 8 of the 31 defects. The Rust build
caught the most. Three planted dependencies between Rust packages create
a dependency cycle, so the build fails, and a broken rename and a use of
the standard library in a crate that forbids it do not compile. Tock's
own crate settings also forbid \texttt{unsafe} code, which Semgrep's
generic Rust rule flags as well. Pylint detected all four broken calls
in one Python pull request, where CA reported one, and SonarQube
detected one of the four. A build of the Java project, which we did not
run, would also have caught the broken rename in that project. Where a
project's rules are written into its build settings, the build enforces
them.

The remaining 23 defects were detected only by CA. They break
project-specific conventions and layering rules in C++, Python,
TypeScript and Java, or a requirement that a new component ship with its
test. No free tool contains these rules by default, and SonarQube
Community does not analyse C, C++ or Rust at all. CA was configured with
each project's rules, including the rule types we planted, so this
comparison shows the value of knowing a project's rules, not that CA is
a better general-purpose analyser. Several of the free tools, such as
Semgrep and SonarQube, accept custom rules, but someone must first write
them for each project.

\section{Threats to validity}\label{sec:threats}

\textbf{Internal validity.} Real projects' rules
may be less complete, which would reduce the benefit of findings. The
scorer was implemented separately from the checker, but by the same
team, so it is not independent in the strict sense, and its agreement
with the checker (kappa 0.84) shows consistency, not correctness. No
blinded audit of final diffs was done. The Alone and +CA workflows
differ in more than the findings. They used different instructions,
because each reflects how the agent would be used in practice, and +CA
allowed a second session when findings remained. The Copilot workflow
used GitHub's own model for the review and a single fix session, unlike
the two sessions allowed with CA, and was run with three repetitions on
24 of the 25 defective pull requests.

\textbf{Construct validity.} A run counts as repaired when a pattern-level specification for each planted defect is
satisfied and every non-test file the pull request added survives and still contains code, under the rules of Section
\ref{sec:outcomes}. This static measure does not by itself show that the code compiles or that the feature still
works, so we added a build and feature check (Section \ref{sec:buildcheck}) and use the build-checked repair rate as
the headline measure. All the main comparisons hold under it, and we report both measures. Every Java, TypeScript and Python repair compiled. Of the 851 repairs, 75 did not pass:
compile failures in Rust and C++, and Python repairs that removed a validator in place, mostly from the smallest
model. The check built 24 of the 25 defective pull requests, not navigation2, and did not run project tests, so
repairs that compile may still be semantically wrong. Four pull-request heads do not compile on their own, and in one
project the added file is not part of the build, which limits what the check can show for those pull requests.

\textbf{External validity.} We studied one model family and one agent.
The free tools in Section \ref{sec:ablation} ran without
project-specific rules, and clang-tidy without the projects' build
configuration, which limits what it can analyse. The pull requests were
relatively compact, with one to three defects of moderate size, 
in changes written to
resemble real contributions. Ten projects in five languages limit
generality to some extent.

\textbf{Conclusion validity.} With 25 defective pull requests, variation
between pull requests is substantial, as the cluster-bootstrap intervals
show. The main comparisons remain significant under tests that treat the
pull request as the unit, but the deletion-check ablation 
is significant for Haiku and Sonnet and not for Fable (Section \ref{sec:deletion}). 
The run-level McNemar tests reported in the tables
treat repetitions as independent and overstate significance. Model
behaviour changes between versions, and our results apply to the
versions named, as of November 2026.

\textbf{Conflict of interest.} The author is affiliated with 
CodeAssure Labs, which develops CodeAssure, the checker 
evaluated in this study. To reduce the risk of bias, 
the defects and their scoring specifications were
 written before any agent run, the scorer does not use the 
 checker, every run is reported, and the outcome of every run and 
the analysis code are public (see Data availability).

\section{Conclusion}\label{sec:conclusion}

In our study, a workflow with external findings that locate each rule
violation and state a repair goal, from one commercial checker,
raised the repair rate of three LLM coding agents on 
planted defects by 43 to 48 percentage points, to 
between 69\% and 93\%, counting only repairs that compile and keep
their feature, with every Java, TypeScript and Python repair
compiling (47 to 58 points, to between 86\% and 100\%, under the
static scorer alone). It also let a
mid-size model outperform the largest model at a fraction of the cost.
Agents often removed a defect by deleting a file the pull request added,
which naive success metrics count as a fix. CA's deletion check 
largely prevents this: in an ablation with the check turned off, 
removals rose from 7 to 19, 2 to 29 and 0 to 17 of 125 runs
for Haiku, Sonnet and Fable, and the repair rate 
fell by 8 to 21 points, a drop significant at the pull-request 
level for Haiku and Sonnet. This workflow also outperformed GitHub 
Copilot code review in its default configuration, and nine free
static-analysis tools run out of the box detected together only 8 of the
31 defects. We recommend that evaluations of coding agents score
deletion separately, and that feedback loops treat it as a finding.

\section*{Acknowledgements}\label{acknowledgements}
\addcontentsline{toc}{section}{Acknowledgements}

This research was supported by the Singapore Ministry of Education (MOE)
Decentralised Gap Funding (MDGF), administered by the Singapore Management
University Institute of Innovation and Entrepreneurship (IIE), Project
No.~MDGF-RIE2025-POC-06. The author gratefully acknowledges this funding.

\section*{Data availability}\label{data-availability}
\addcontentsline{toc}{section}{Data availability}

The outcome of every run (an opaque identifier for the pull request,
defect type, model, workflow, repetition, scorer outcome, sessions,
steps and total cost) the detection result of every tool for each
planted defect, and a script that recomputes every table and test in
this paper are public at
\url{https://github.com/CodeAssureLabs/external-feedback-coding-agents}.
They allow the statistics to be checked, but not the runs to be
re-scored. Full per-run records (checker findings, agent transcripts and
diffs), the scorer specifications and the exact agent instructions are
available   on request under a confidentiality agreement,
from hello@codeassurelabs.com. The checker's rule format and internal
components are not described here.

\bibliography{references.bib}

\end{document}